# Potential of longevity: hidden in structural complexity

Jicun Wang-Michelitsch[1]*, Thomas M Michelitsch[2]

[1]Department of Medicine, Addenbrooke's Hospital, University Cambridge, UK (work address until 2007)

[2]Institut Jean le Rond d'Alembert (Paris 6), CNRS UMR 7190 Paris, France

**Abstract**

In order to understand the phenomenon of longevity in biological world, the relationship between the potential of longevity and the structural complexity of an organism is analyzed. **I.** The potential of longevity is the maximum lifespan of an organism if it lives in an "ideal" environment. The longevity of an organism includes two parts: the time for development (development time) and the time for structure-maintenance (maintenance time). **II.** The development time of an organism depends on its structural complexity. The maintenance time is related to two factors: the degree of damage-exposure and the potential of functionality for structure-maintenance. The potential of functionality includes also two parts: the capacity of basic functionality and the potential of functional compensation. The capacity of basic functionality is built in the structural complexity of an organism. For an organism, the structural complexity and the functionality will be reduced gradually with age by accumulation of Misrepairs. However, functional compensation can slow down the decline of functionality with aging. There are two major mechanisms for functional compensation: network-like organization of sub-structures and regeneration of sub-structures. These two mechanisms are also built in the structure of an organism. Since the development time and the maintenance time are both determined by structural complexity, the potential of longevity of an organism is hidden in structural complexity. **III.** The individuals of different species' have different longevities because they have different structural complexities. An animal has limited longevity because it has limited structural complexity. Limited structural complexity and limited longevity are essential for the survival of a species. **IV.** Despite having the same potential of longevity, the individuals of a species can have different lifespans. The lifespan of an individual is more related to the degree of damage-exposure, which is determined by the living circumstance and the living habits of the individual. **In conclusion,** the potential of longevity of an organism is hidden in structural complexity, but the real lifespan of an organism is more related to the living environment.

## Keywords

To have a long lifespan is a dream for all of us. However, to realize this dream, we need to know at first what factors determine our longevity and why we all have a limited longevity. Some research groups are searching for "lifespan-related genes" for uncovering the secret of longevity. However, lifespan-related genes are not necessarily lifespan-determined. Aging is related to longevity; however a concrete link between aging and longevity is missing. For interpreting aging, we have proposed a novel theory: Misrepair-accumulation theory (Wang, 2009). The main idea of this theory is: aging of a multi-cellular organism is a result of accumulation of Misrepairs on tissue level. This theory is helpful not only for understanding aging but also for understanding longevity. In the present paper, we discuss the relationship between the longevity and the structural complexity of an organism by Misrepair mechanism. We aim to show by our discussion that the potential of longevity of an organism is determined by its structural complexity. Our discussion tackles the following issues:

I.   Concepts of longevity

II.  The potential of longevity of an organism is hidden in structural complexity

  2.1   Concept of structural complexity of an organism
       2.1.1    Determining factors for the structural complexity of an organism
       2.1.2    The time for development of an organism is determined by the degree of structural complexity
  2.2   Gradual reduction of structural complexity of an organism with aging: by accumulation of Misrepairs
  2.3   The potential of functionality of an organism for structure-maintenance is built in structural complexity
       2.3.1    Basic functionality: related to the size of an organism
       2.3.2    Functional compensation by network-like organization of sub-structures
       2.3.3    Functional compensation by regeneration of sub-structures
  2.4   The potential of longevity of an organism is determined by structural complexity
  2.5   Limited potential of longevity of an animal is a result of limited structural complexity
  2.6   The long longevity of a queen ant is obtained by redirected development
  2.7   The long longevity of a tree is obtained by repeated developments

III. Individual lifespans: more related to living environments and living habits

IV.  Conclusions

## I. Concepts of longevity

The term of "longevity" refers to the life expectancy of an organism, namely the length of time that an organism is expected to exist as a whole structure. It includes two parts: the time for development of the organism and the remaining time used for maintaining the structure of the organism till breakdown (death) of the structure. Before discussing longevity, several concepts related to longevity need to be clearly defined and distinguished. These concepts include lifespan, average longevity, and potential of longevity. "Longevity" is a statistical prediction of life-length, and it is strictly speaking different from the actual life-length of an organism, which is called "lifespan". The longevity of a group of individuals of a species is often predicted by the average life-length of individuals, which is called average longevity. The individuals that are of the same species but live in different geographic areas can have quite different average longevities.

However, when we compare the longevities of different species', we are actually studying their maximum longevities. The maximum longevity of individuals of a species is in fact the "potential of longevity" of an individual. The potential of longevity is the life expectancy of an organism when it lives in an "ideal" environment. "Ideal" environment maybe does not exist in nature; however an environment where an organism can obtain sufficient food and be exposed to the least damage is regarded as an "ideal" environment. In this paper, three longevity-related concepts have their definitions as follows:

- Lifespan: the actual life-length of an organism
- Average longevity: the average life expectancy of a group of individuals living in similar environments
- Potential of longevity: the maximum life expectancy of individuals of a species living in an ideal environment

## II. The potential of longevity of an organism is hidden in structural complexity

Two factors are related to the potential of longevity of an organism: the complexity of its structure and the potential of its functionality. The complexity of structure determines the time for development of the organism. The potential of functionality determines the time for structure-maintenance till death of the organism. It is known that the functionality of a system is determined by the structure of the system. Thus, the potential of functionality of an organism should lie in the complexity of the structure of the organism. In this part, we discuss how the functionality and the potential of longevity of an organism are determined by the complexity of structure.

### 2.1 Concept of structural complexity of an organism

In physics, "complexity" is a term used for describing the relationship between the whole system and its sub-systems in a complex system such as a living being. A complex system manifests its complexity on several aspects, including emergence, feedback effect, self-organization, and adaptation (Haken, 1990; Wunderlin, 1992). **Emergence** is the phenomenon that the behavior of a system cannot be tracked back to the behaviors of its sub-systems. For example, a heart has the ability of pumping blood, but a singular cardiac muscular cell, as part of heart wall, does not have this ability. **Feedback effect** is the phenomenon that the sub-systems can make responses to changes of the whole system, and wise-verse. For example, cardiac muscular cells are the functional components of a heart; however failure of heart caused by arterial hypertension may result in death of cardiac cells. **Self-organization** is the phenomenon that a system can develop its structure automatically by organizing its sub-systems given sufficient energy and substances. Development of embryo is an example of self-organization. **Adaptation** is the phenomenon that the sub-systems can make suitable responses to changes of environment for preventing death of the whole system. For example, overloading of heart by hypertension will induce enlargement of cardiac cells, which can make functional compensation for the heart.

All of these phenomena show the complex functional relationships between the whole system and its sub-systems. However, all of these functional relationships are built in the organization of sub-systems, namely the special spatial relationship of sub-systems. For a system or an organism, functional complexity is built in **structural complexity**. Therefore, for understanding the potential of functionality of a system, it is essential to study the **structural complexity** of the system. In this paper, **structural complexity** is defined as the complexity of organization of the sub-systems (or sub-structures) of a system (or an organism).

### 2.1.1 Determining factors for the structural complexity of an organism

The degree of structural complexity of an organism is related to several factors. Firstly, it is in ratio to the total number of sub-structures, including the number of hierarchical levels of sub-structures, the number of types of sub-structures, and the amount of sub-structures of each type. For example, molecules, cells, tissues, and organs are the four levels of sub-structures of a multi-cellular organism. In a tissue, there are different types of cells and extracellular matrixes (ECMs). The structural complexity of a tissue will increase with the increase of diversity of cells/ECMs and the increase of total amounts of cells/ECMs. Secondly, the manner of organization (distribution) of sub-structures also contributes to the structural complexity of an organism. An organization will have a higher complexity if it permits each sub-structure to communicate with more other sub-structures. As shown in Figure 1, given the same total number of sub-structures, organization **C** has higher complexity than organization **A** and organization **B**. In organization **C,** each sub-structure (such as sub-structure **Z**) can communicate with six other sub-structures, whereas in organization **A** (or organization **B**), each sub-structure can communicate only with three (or four) other sub-structures.

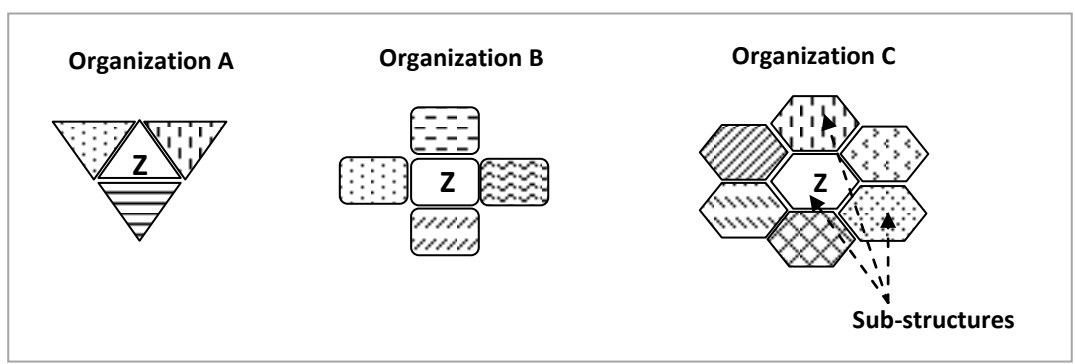

Figure 1. Different structural complexities in different manners of organizations of sub-structures

An organization will have higher complexity when it permits each sub-structure to communicate with more other sub-structures. For example, organization **C** has higher complexity than organization **A** and organization **B**. In organization **C,** each sub-structure (such as sub-structure **Z**) can communicate with six other sub-structures. In organization **A** (or organization **B**), each sub-structure can communicate only with three (or four) other sub-structures.

The multiple communicating pathways between sub-structures compose a communicating network. For example, the lobules of liver have a network-like organization (Figure 2). In a liver, every lobule has six branches of portal veins, and three neighbor lobules share a branch of portal vein. By such an organization, each lobule functions as a common pathway for six branches of portal veins to "be linked to" a central vein; and each portal vein has three pathways to "communicate" with central veins (via three lobules). The organization of lobules in this way reduces the risk of failure of liver when some lobules or portal veins fail. Thus, network-like organization of cells/tissues is an effective way to make functional compensation for an organ. The degree of complexity of a network is related to two factors: the number of sub-structures (points) and the number of communicating pathways of each sub-structure with other sub-structures. For example, the degree of complexity of the neuron network in brain is related to the number of neurons and the number of dendrites of each neuron.

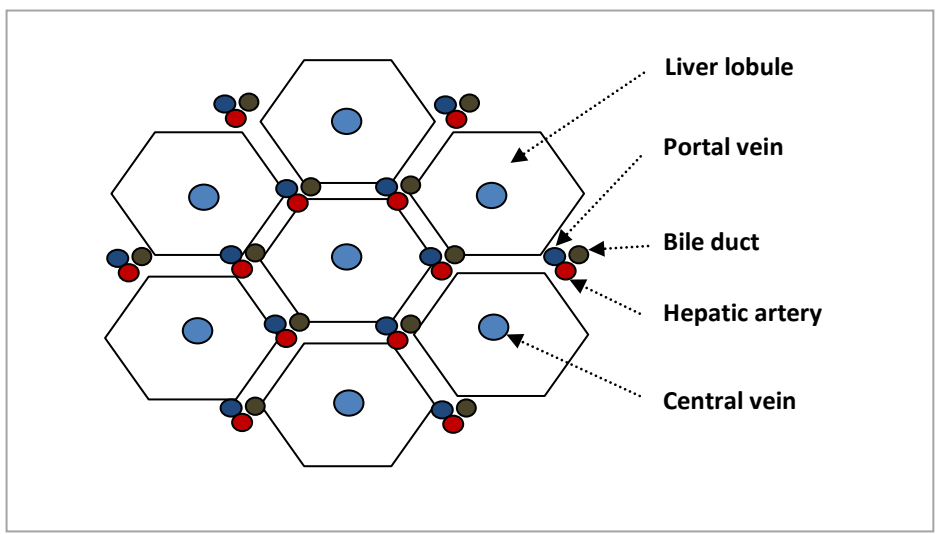

**Figure 2. Network-like organization of lobules in liver**

In a liver, the lobules are in a network-like organization. Every lobule has six branches of portal veins, and three neighbor lobules share a branch of portal vein. By such an organization, each lobule functions as a common pathway for six branches of portal veins to be connected to a central vein, and each portal vein has three pathways to "communicate" with central veins (via three lobules). The network-like organization of lobules is an effective way to reduce the risk of failure of liver when some lobules or portal veins fail.

*email : thomasjicun@gmail.com                                                                                                                 5

In summary, the degree of structural complexity of an organism is not only related to the number of levels of sub-structures and the diversity/amount of sub-structures, but also related to the number of communicating pathways between sub-structures (Box 1). The total number of sub-structures may be the main determining factor for the size of an organism (or organ). However, the degree of complexity of communicating network of sub-structures is not necessarily related to the size of an organism (or organ). Therefore, the degree of structural complexity is partially related to the size of an organism. For example, a rat has a bigger body than a mouse, thus the structural complexity of a rat is higher than that of a mouse. However, when we compare the brain of a human being with that of a cow, the result is not obvious. A human brain is smaller than a cow brain, but the network of neurons in human brain is more complex than that of a cow. Therefore, a human brain has higher structural complexity than a cow brain.

**Box 1. Determining factors for the structural complexity of an organism**

- The number of levels of sub-structures
- The diversity and amount of sub-structures at each level
- The number of communicating pathways between sub-structures

### 2.1.2 The time for development of an organism is determined by the degree of structural complexity

Development of an organism is a process of construction of a structure and its complexity. A more complex organism needs a longer time for development. The animals of different species' have different sizes of organs; therefore they have different degrees of structural complexity and need different lengths of time for development. However, the period of development of an organism cannot be too long. If it is too long, the organism could die before being fully developed. In nature, full development of an organism is only possible when the rate of construction is higher than that of destruction by damage. An embryo needs to develop in a protected environment, since the high structural complexity of an animal embryo cannot be built up successfully in a damaging natural environment. Natural pressure restricts the increase of structural complexity of creatures. An animal has limited structural complexity because it has limited body development. Different species' of animals have different limits on structural complexity, and the limits are determined by their gene configurations. Differently, some plants including most species' of trees undergo repeated developments after reproduction age. The repeated developments increase continuously the structural complexities of a tree. Some trees seem to have no genetic limit on structural complexity; however their developments will be stopped by a catastrophe in nature.

### 2.2 Gradual reduction of structural complexity of an organism with aging: by accumulation of Misrepairs

An organism is able to maintain its structure by exerting functions on many aspects. The potential of functionality of an organism is built in its structure, which is normally fully developed. An organism goes to failure on functionality in two ways. One is rapid, directly by a severe injury, which destroys the structural integrity of an organism. Rapid death is an accidental death. The other is gradual, through a process of aging (Figure 3). In our view, aging of an organism is a result of accumulation of Misrepairs of its structure (Wang, 2009). Misrepair is a strategy of repair of an injured living structure by altered materials and in altered remodeling. In situation of a severe injury, when full repair is impossible to achieve, Misrepair is a way to maintain the structural integrity and prevent death of an organism. Misrepairs are unavoidable for an individual to survive till the age of reproduction. Therefore, Misrepair mechanism is essential for the survival of a species.

However, Misrepair results in structure-alteration and function-reduction of a living structure. Scar formation is a kind of Misrepair essential for healing of a deep wound. However, a skin scar alters the structure/function of the local part of skin. Misrepairs are irreversible and irremovable, thus they can accumulate with time, appearing as aging of a structure. Accumulation of Misrepairs results in a gradual reduction of structural complexity and functionality of an organism, till death of the organism by failure of functions. Therefore, aging is a result of long-term struggling of an organism with damaging environment. The more destructive the environment is, the more rapid will an organism lose its structural complexity and functionality.

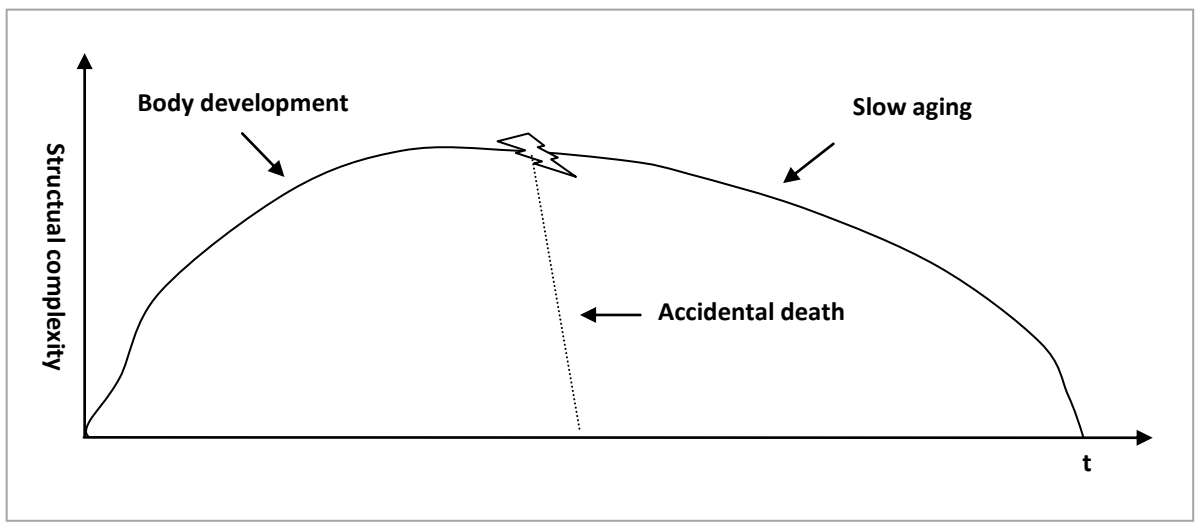

**Figure 3. Gradual reduction of structural complexity of an organism with aging**

An organism goes to failure on functionality in two ways. One is rapid, directly by a severe injury, which destroys the structural integrity of an organism. Rapid death is an accidental death. The other is gradual, through a process of aging. In our view, aging of an organism is a result of accumulation of Misrepairs of its structure. Misrepair results in irreversible structure-alteration and function-reduction of a living structure. Thus, accumulation of Misrepairs leads to gradual reduction of structural complexity of an organism with age, till death of the organism by failure of functions.

## 2.3 The potential of functionality of an organism for structure-maintenance is built in structural complexity

An organism can exert many functions to maintain its structure. The potential of functionality of an organism includes two parts: the capacity of basic functionality and the potential of functional compensation. The capacity of basic functionality is related to the total number of sub-structures, thus it is built in the structural complexity of an organism. Functional compensation refers to the capability of an organism on making up part of lost functionality when a sub-structure is dead. Functional compensation can slow down the decline of functionality of an organism with aging. There are at least two mechanisms for functional compensation: network-like organization of sub-structures and regeneration of sub-structures. The potential of functional compensation by these two mechanisms are also built in the structural complexity of an organism.

### *2.3.1 Basic functionality: related to the size of an organism*

The basic functionality of an organism (or organ) can be approximately evaluated by the size of the organism (or organ). The size of an organism is in ratio to the total number of sub-structures. For example, the digesting capacity of a stomach is determined by the number of cells and glands in stomach mucosa. The sizes of the same type of organs are different in the animals of different species. Thus the hearts/livers in different species of animals have different functional potentials. For example, the heart of a rat has higher functional potential than that of a mouse, because the former is bigger and the heart wall of a rat is thicker than that of a mouse. However, for all animals, the basic functionality will be reduced with aging by accumulation of Misrepairs and gradual loss of functional unites. After age 60, the functionality of a human brain, including the ability of memorizing, decreases gradually. The reason for that is: an old people will have less and less number of functional neurons in brain with aging. For the same person, the brain is smaller at age 80 than that at age 30. Accumulation of Misrepairs reduces the number of functional sub-structures of an organism with aging. Nevertheless, a bigger organism has often bigger organs, thus it needs more time and more number of Misrepairs than a smaller one for going to failure on functions.

### *2.3.2 Functional compensation by network-like organization of sub-structures*

A network-like organization of sub-structures permits functional substitution of sub-structures. The lobules of liver are in a network-like organization. In a network, the communications between two sub-structures (points) have more than one pathway, and each sub-structure (point) functions as a crossing point for several communicating pathways between other sub-structures. When there are multiple pathways between two points (point A and point B) by "other points (such as point C or point D)", these pathways can substitute to each other. Namely, point C and point D can substitute each other on mediating the communications between point A and point B. Even if point C fails, point D can maintain the functionality of the whole network. For example, in a network like that in Figure 4A, sub-structure **X2** can communicate with sub-structure **Z2** via two pathways: **Y1** and **Y2**. **Y1** can communicate with **Y2** via pathway **X2** and pathway **Z2**. The pathway between **X2** and **Z1** and the pathway between **X1** and **Z2** cross at **Y1**. In mediating the communications between **X2** and **Z2**, **Y1** can function as a substitution for **Y2**. When **Y1** fails, the function of **Y1** can be substituted by **Y2**; and the communicating efficiency between **X2** and **Z2** will not be severely affected.

**A**

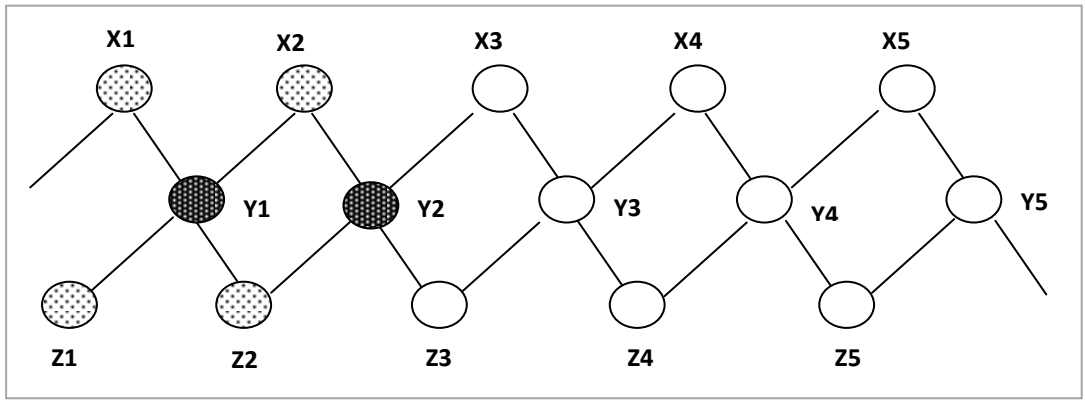

**B**

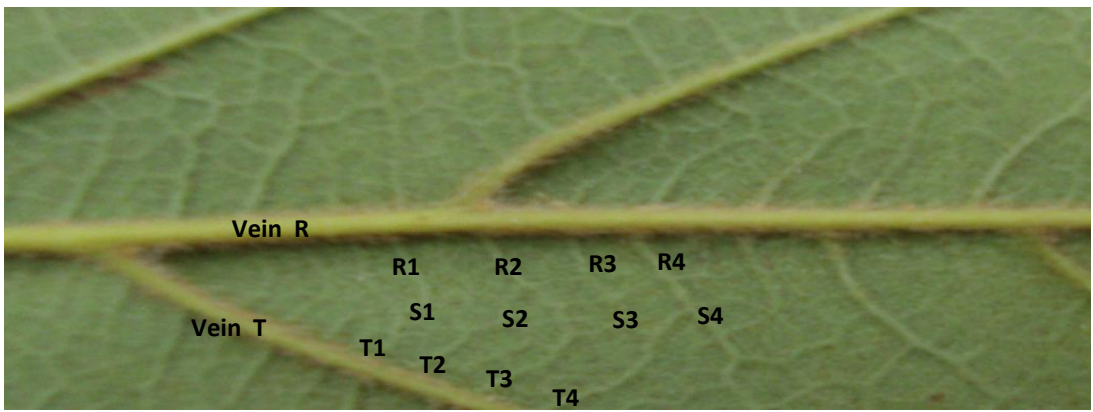

**Figure 4. Functional compensation of an organism by network-like organization of sub-structures**

A. **Communicating network between sub-structures**  In the organization composed by sub-structures **X1, X2**..., **Y1, Y2**..., and **Z1, Z2**..., sub-structure **X2** can communicate with sub-structure **Z2** via two pathways: **Y1** and **Y2**. **Y1** can communicate with **Y2** via pathway **X2** and pathway **Z2**. The pathway between **X2** and **Z1** and the pathway between **X1** and **Z2** cross at **Y1**. These pathways between different sub-structures cross with each other and compose a network. In mediating the communications between **X2** and **Z2**, **Y1** can function as a substitution for **Y2**. When **Y1** fails, the function of **Y1** can be substituted by **Y2**; and the communicating efficiency between **X2** and **Z2** will not be severely affected.

B. **Network-like distribution of veins in a plant leaf**  In the leaf, vein R1 crosses with vein T1 at point S1. This means that the part of leaf at point S1 has two sources of water supply: vein R1 and vein T1. For the water supply of point S1, vein T1 can substitute vein R1. Similarly, the parts of leaf at points S2, S3, and S4 have all two sources of water supply, namely vein T and vein R. The network-like distribution of branches of veins reduces the risk of death of a leaf from failure of a branch of vein.

The distribution of small blood vessels in an organ is also in a network, because some branches of vessels cross to each other. This network is similar to that of veins in a plant leaf. In the leaf in Figure 4B, vein **R1** crosses vein **T1** at point **S1**. This means that the part of tissue at point **S1** has two sources of water supply: vein R1 and vein **T1**. For the water supply to **S1**, vein **T1** can substitute vein **R1**. Similarly, the parts of tissue at points **S2, S3,** and **S4** have all two sources of water supply: vein **T** and vein **R**. A network-like distribution of branches of blood vessels reduces the risk of death of an organ from blockage of a branch of blood vessel. The potential of functional compensation by network-like organization of sub-structures is related to the degree of complexity of network, thus it is determined by the structural complexity of an organism.

### 2.3.3 Functional compensation by regeneration of sub-structures

A living organism has the ability to reproduce sub-structures for maintaining its structure and functionality. These sub-structures include cells, proteins, lipids, and small molecules. For a multi-cellular organism, reproduction of cells is essential for the repair/maintenance of the organism. Except in hematopoietic tissue and epithelial tissue, regeneration of cells is often induced by death of cells, and new cells are used for replacing the dead ones. Regeneration of cells has two roles for an organ: to maintain the structural integrity and to compensate the lost functionality due to loss of cells. A tissue may have reduced functionality when some cells die. By making up structural complexity, regeneration of cells can compensate the functionality of a tissue, completely by full repair or partially by Misrepair (Figure 5). Misrepair of a tissue is often achieved by regeneration of cells. However, in a Misrepair, the reproduced cells are in an altered reorganization. Thus the cell regeneration in Misrepair can only partially compensate the lost functionality of a tissue (Figure 5). Regeneration of cells can slow down the gradual reduction of functionality of an organism with aging. The potential of functional compensation by cell-regeneration is related to the number of stem cells, thus it is built in the structural complexity of an organism.

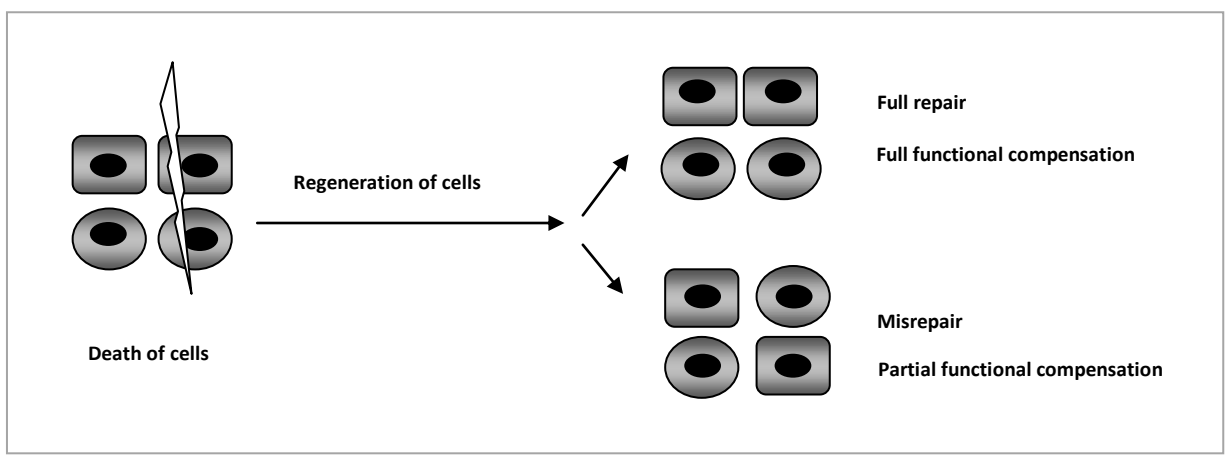

**Figure 5. Functional compensation by regeneration of cells in full repair and in Misrepair**

Regeneration of cells in a tissue is often induced by death of cells, and new cells are used for replacing the dead ones. Full repair of a tissue is often accomplished by regeneration of cells. In full repair, the structural complexity and the functionality of a tissue is completely restored by regenerated cells (**full repair**). Misrepair can be also achieved by regeneration of cells; however the reproduced cells in Misrepair are in an altered reorganization. Thus the cell regeneration in Misrepair can only partially compensate the lost functionality of a tissue (**Misrepair**).

Taken together, the potential of functional compensation by network-like organization of sub-structures and that by regeneration of sub-structures are both built in the structural complexity of an organism. Small creatures such as worms have often short longevity, because they have low structural complexity. With low structural complexity, small animals have low basic functionality and low potential of functional compensation for structure-maintenance. Normally, basic functionality is related to the size of an organism, but the potential of functional compensation is not necessarily related. Therefore, the total potential of functionality is not completely determined by the size of an organism. For example, the potential of functionality of the brain of an animal cannot be judged by the size of brain. A cow has a much bigger brain that a mouse; however a cow is not many times "cleverer" than a mouse. In summary, the potential of functionality of an organism for structure-maintenance is built in the structural complexity of the organism.

## 2.4 The potential of longevity of an organism is determined by structural complexity

The time for body development and the time for structure-maintenance are both related to the structural complexity of an organism. On one hand, an organism having higher structural complexity needs a longer time to build up. On the other hand, a higher structural complexity gives a higher potential of repair/maintenance of an organism and permits a longer period of time of structure-maintenance. Taken together, a longer development time plus a longer maintenance time makes a longer potential of longevity of an organism (Figure 6).

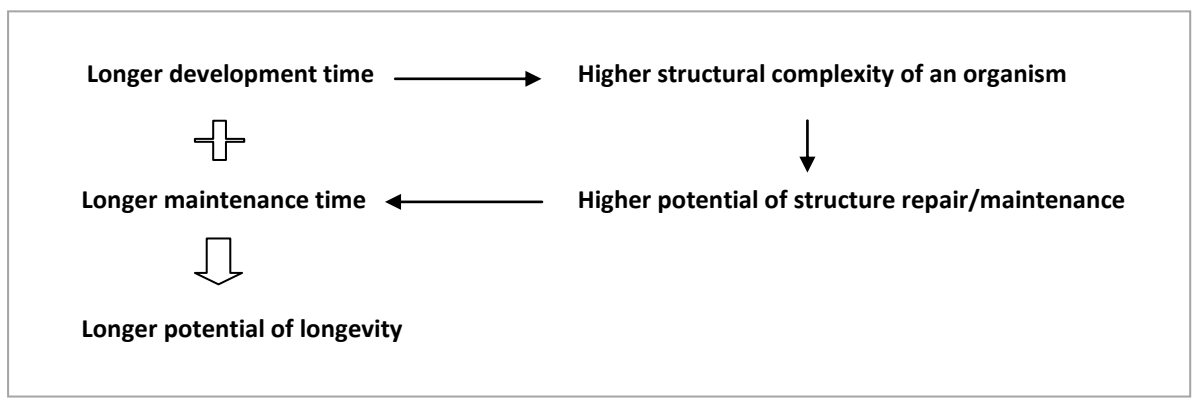

**Figure 6. The potential of longevity of an organism is hidden in structural complexity**

The time for development and the time for structure-maintenance are both related to the structural complexity of an organism. On one hand, an organism having higher structural complexity needs a longer time to build up. On the other hand, a higher structural complexity gives a higher potential of repair/maintenance and permits a longer period of structure-maintenance. Taken together, a longer development time plus a longer maintenance time makes a longer potential of longevity of an organism.

The individuals of different species' have different potentials of longevity because they have different structural complexities. In the same living environment, the individuals of a species that have higher structural complexity will have longer potential of longevity. As a human being, we have bigger organs than a mouse. Thus, our body has higher structural complexity than that of a mouse. Therefore, we need a longer time for body development and we can maintain our body for a longer period of time than a mouse. For example, human heart is bigger than that of a mouse, thus the lumens of the main cardiac arteries are larger in human heart than in mouse heart. It is known that atherosclerosis is the main causing factor for blockage of cardiac arteries and failure of heart in animals. However, for blocking a cardiac artery in mouse heart, a small atherosclerotic plaque is sufficient. But for a human, only a big atherosclerotic plaque can cause failure of heart. Therefore, human heart and human cardiac arteries have higher functional potential than that of a mouse, and it takes longer time for human heart to go to failure.

## 2.5 Limited potential of longevity of an animal is a result of limited structural complexity

An animal has limited potential of longevity, because it has limited structural complexity. Limited structural complexity gives a limited potential of repair/maintenance for an animal. For example, the diameters of the main cardiac arteries in a human are fixed. The non-stop growth of atherosclerotic plaques with time will finally lead to blockage of an artery by a plaque. The fixed body structure of an adult animal is determined by the gene configuration of the species. It is possible that new species' of animals that have higher structural complexities than existed animals appear by evolution. However, the increase of structural complexity of creatures will be restricted by the destructive pressure of nature. Large animals including whales and elephants have often low populations. One reason is that: these animals have big bodies and they need a long time for body development; thus many children whales/elephants cannot survive till reproduction age in natural environment. If a species of animal is too complex on body structure and most individuals die before reproduction age, the species will die out. Therefore, limited structural complexity and limited longevity are essential for the survival of a species.

## 2.6 The long longevity of a queen ant is obtained by redirected development

The big difference between a queen ant and a worker ant on lifespan is amazing. Despite having the same gene configuration, a queen ant develops in a different way from other female ants. A queen ant continues the body development and the body growth after age of maturation. However, for other female ants, the body development stops at age of maturation. The redirected development of queen ant is induced by environment factors. This altered development makes a queen ant have a distinct body structure from a worker ant. Namely, the redirected development makes a queen ant have a higher structural complexity, a higher potential of functionality, and a longer potential of longevity than other ants (Figure 7). Apart from higher structural complexity, living in a protected environment may also contribute to the longer lifespan of a queen ant.

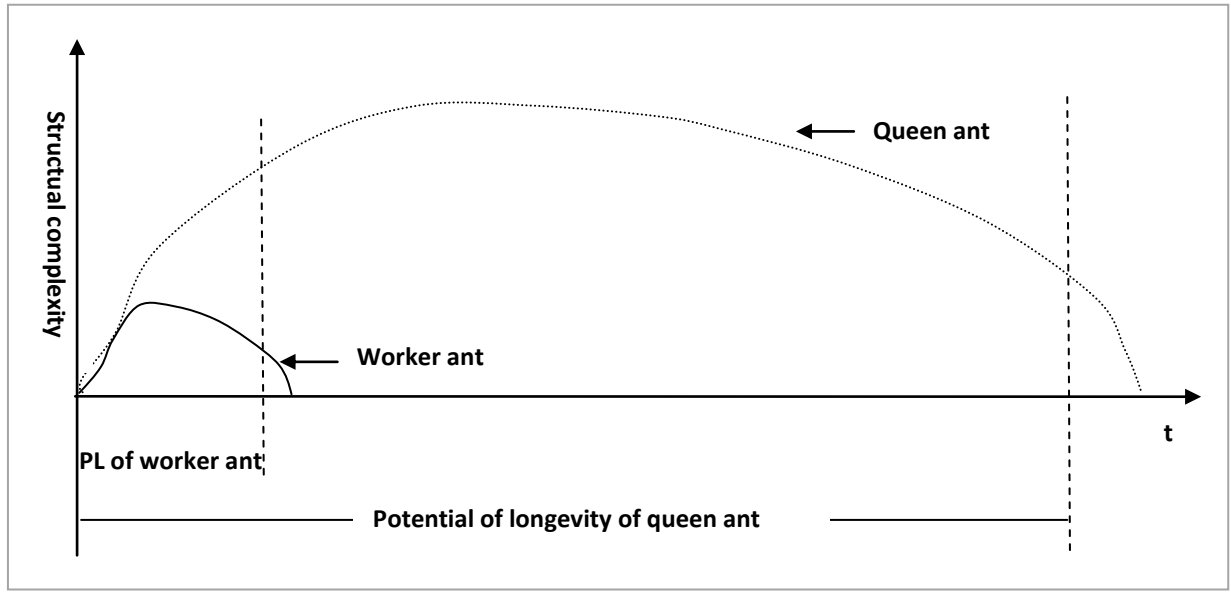

**Figure 7. Differences on development and on longevity between a queen ant and a worker ant**

Despite having the same gene configuration, a queen ant develops differently from other female ants. The difference on development between them leads to their differences on the potential of longevity (**PL**). A queen ant can continue developing and growing after age of maturation (**Queen ant**). Other female ants stop development at age of maturation (**Worker ant**). The redirected development of a queen ant makes her have a higher structural complexity, a higher potential of functionality (for structure-maintenance), and a longer potential of longevity than other ants.

It is found that some genes are "aging-related" or "lifespan-related". For some scientists, gene modification is a strategy for extending the longevity of animals. However, if gene modification could extend the lifespans of animals, the effect should be made by altering the process of body development rather than by retarding aging. Modification of a gene can possibly redirect body development and alter the final structure of an organism. With altered structural complexity and altered potential of functionality, an organism may have an altered potential of longevity. In one study, the research group of Helfand has successfully extended the lifespan of *drosophila* two times by introducing a mutation on gene *Ingy* (Mardon, 2003). Although the mutant *drosophilae* have normal metabolism and normal flying ability, the reproducing ability of these individuals was reduced significantly, especially in a condition of low calorie nutrition. In another study, the mice that have lower expression of protein mTOR have longer lifespans (Wu, 2013). The mutant mice are healthy; however they are slightly smaller than normal mice and are more sensitive to infections.

In these studies, although it is unknown how a genetic modification affects body development, alteration of development is evident. The changes on body size, immunity, and/or ability of reproduction are all evidences of alteration of body development. Therefore, in our view, a strategy for extending the longevity of animals by gene modification cannot be successful. Such a strategy may lead to defective body development. With defects on functionality, lower chance of survival, and lower rate of reproduction, the modified species cannot survive for a long time.

## 2.7 The long longevity of a tree is obtained by repeated developments

Many species' of trees can survive much longer than animals. One reason is that: trees can develop repeatedly each year after age of maturation, but an animal has only one-time development. By repeated developments, a tree can obtain more and more structural complexity and functionality (Figure 8). Each year, a tree will have new branches, and the main trunk and "old" branches will become thicker. With time, part of the trunk and some branches of a tree may lose their functionality; however new part of the trunk and new branches can compensate the lost functionality. The repeated developments of a tree are genetically controlled. Some trees seem to have no genetic limit on structural complexity; however a catastrophe in nature will terminate the developments of a tree. The lifespan of a tree is finally a result of competition between construction and destruction of the tree in nature.

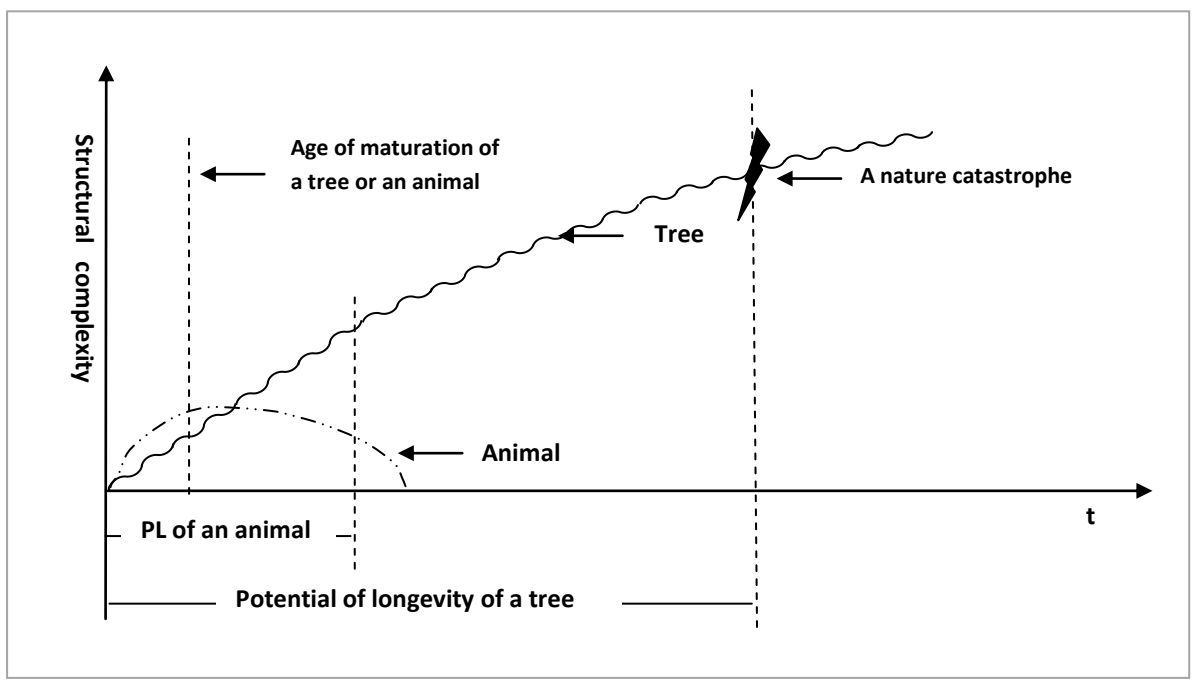

**Figure 8. The long longevity of a tree is obtained by repeated developments**

Some species' of trees have much longer potential of longevity (**PL**) than animals. A main reason is that a tree undergoes repeated developments each year after age of maturation but an animal has only one-time development. By repeated developments, a tree can obtain more and more structural complexity and functionality. Some trees seem to have no genetic limit on structural complexity; however a nature catastrophe can terminate the developments of a tree.

Among all species' of trees, those that have longer potentials of longevity often have higher functionalities on defense and self-protection. These species' of trees have thus higher structural complexities, and need longer time for body development till reproduction age. For example, a Ginkgo biloba tree can have longevity of more than 3000 years; but it needs 20 years for maturation. A willow tree has longevity of 150 years; but it needs only 2-3 years for maturation. Thus, a tree that has longer potential of longevity will have higher risk of death before age of maturation. The species' of trees that have extreme long potentials of longevity such as the Ginkgo biloba trees are often the species' that are on the edge of extinction.

### III. Individual lifespans: more related to living environments and living habits

Despite having the same potential of longevity, the individuals of a species can have quite different lifespans. Twin brothers have often different lifespans although they have the same genetic background and a similar structural complexity. Therefore, the lifespan of an organism is not completely determined by structure complexity. In this part, we will discuss the aging-related lifespan of human being. With age, we are all approaching to death by more and more diseases, such as tumors, arterial hypertension, and atherosclerosis. The direct cause of death of our body is often the failure of a key organ such as the heart and the brain. Failure of a key organ can be acute, progressive, or chronic.

**Acute failure of an organ** is often a consequence of breakdown of structural integrity of the organ by a severe injury. However, breakdown of an organ is often caused by aging of this organ or other organs. For example, cerebral bleeding is a fatal disease because it can destroy the structural integrity of the brain. However, cerebral bleeding is often a consequence of arterial hypertension due to aging of arterial walls. **Progressive failure of an organ** is often a consequence of tumor occupation to the organ. For example, a patient with colon cancer dies often from failure of liver, because colon cancer cells can invade into liver via portal veins. **Chronic failure of an organ** is a consequence of gradual loss of functionality due to aging of the organ. Alzheimer syndrome is a disease caused by aging and chronic failure of the brain.

Three factors may affect the "lifespan" of a key organ: rate of aging of the organ, inducing factor of failure of the organ, and random factors. Firstly, **the rate of aging of an organ** is determined by the degree and the frequency of damage-exposure of the organ. Thus, the rate of aging of an organ is closely related to the living circumstance and the living habit of an individual. The rates of aging of tissues can be quite different in different organs and in different individuals. For example, smokers may have much quicker aging of lung than non-smokers. The individuals that are often exposed to strong sunlight may have accelerated aging of skin. The individuals who live in a polluted environment may have increased risk of chronic pulmonary inflammations and accelerated aging of lung.

Secondly, complete failure of an organ takes place often when there is an **inducing factor**. Inducing factors are the factors that can cause overload of an aged organ. For example, arterial hypertension is a main causing factor for cerebral bleeding; however occurrence of cerebral bleeding is often promoted by a strong activation of sympathetic nerves and the heart. Excitement of sympathetic nerves can increase further the blood pressure to artery walls and increase the risk of disruption of an artery in brain. Cancer patients die often from infections, because the immune system of a cancer patient can be severely destroyed by the invasive cancer cells. Thirdly, some **random factors** are also related to the lifespan of an organ. For example, the degree of malignancy of a tumor determines how long an affected organ can "survive". However, the grade of malignancy of a tumor seems to be unpredictable. Taken together, the lifespan of a key organ is related to multiple factors, and the lifespan of an individual is more determined by the living circumstance and the living habit of the individual. Therefore, for extending our lifespan, the most important is to reduce the risk of damage-exposure.

Some animals gain their long lifespans by their special competences on self-protection in nature. The competence of self-protection is a part of functionality of an organism. A tortoise can survive more than 100 years. However, tortoises obtain their long longevity by long time of hibernation. A tortoise hibernates for half of its lifetime. During hibernation, a tortoise has reduced risk of exposure to damage, such as bad weather and dangerous animals. An african elephant can have a lifespan of 70 years. One reason is that an elephant has almost no natural enemy except human being! For human being, one of our secrets to have a longer longevity than other animals is that we can build up a protective living environment. With higher intelligence, we can organize a society composed of agriculture, industry, military, medical system, and etc. In such a society, we have higher living security. Therefore, it is the civilization that has given us a longer and longer longevity in the last five centuries.

## IV. Conclusions

We have discussed in this paper the determining factors for the potential of longevity of an organism and the influencing factors for the lifespan of an individual. Development of an organism is a process of building-up of structural complexity and functionality. However, in natural environments, the structural complexity of an organism will be gradually reduced with age by accumulation of Misrepairs. The maintenance time of an organism is related to the potential of its functionality, which is built in structural complexity. Thus, the structural complexity of an organism determines not only the time for maturation but also the time for structure-maintenance. Namely, the potential of longevity of an organism is determined by the structural complexity of the organism. For animals, limited longevity is a result of limited structural complexity. Importantly, limited structural complexity and limited longevity are essential for the survival of a species. Some trees have long longevity because they can obtain additional structural complexity by repeated developments. A queen ant has much longer longevity than a worker ant, because the queen ant undergoes a "redirected" development. Despite having the same potential of longevity, the individuals of a species can have different lifespans. The lifespan of an individual is more related to his living circumstance and his living habit.